\documentclass[twocolumn,twocolappendix]{aastex701}
\usepackage{CJK}

\graphicspath{{./}{figures/}{../Figure}{../Figure/mock/}}

\begin{document}
\begin{CJK*}{UTF8}{gbsn}

\title{Physical Properties of 6.7 Million Galaxies from the DESI Bright Galaxy Survey: 
       Spectral Fitting and Systematic Tests with Mock Spectra}

\author[0000-0002-0656-075X, gname=Niu, sname=Li]{Niu Li}
\affiliation{National Astronomical Observatories, Chinese Academy of Sciences, Beijing 100101, China}
\email[show]{liniu@nao.cas.cn}

\author[0000-0002-6684-3997, gname=Hu, sname=Zou]{Hu Zou}
\affiliation{National Astronomical Observatories, Chinese Academy of Sciences, Beijing 100101, China}
\affiliation{School of Astronomy and Space Science, University of Chinese Academy of Sciences, Beijing 101408, China}
\email[show]{zouhu@nao.cas.cn}

\author[]{Jinfu Gou}
\affiliation{National Astronomical Observatories, Chinese Academy of Sciences, Beijing 100101, China}
\affiliation{School of Astronomy and Space Science, University of Chinese Academy of Sciences, Beijing 101408, China}
\email[]{xxx@nao.cas.cn}

\author[]{Weijian Guo}
\affiliation{National Astronomical Observatories, Chinese Academy of Sciences, Beijing 100101, China}
\email[]{xxx@nao.cas.cn}

\author[]{Wenxiong Li}
\affiliation{National Astronomical Observatories, Chinese Academy of Sciences, Beijing 100101, China}
\email[]{xxx@nao.cas.cn}

\author[]{Haoming Song}
\affiliation{National Astronomical Observatories, Chinese Academy of Sciences, Beijing 100101, China}
\affiliation{School of Astronomy and Space Science, University of Chinese Academy of Sciences, Beijing 101408, China}
\email[]{xxx@nao.cas.cn}

\author[]{Jipeng Sui}
\affiliation{National Astronomical Observatories, Chinese Academy of Sciences, Beijing 100101, China}
\affiliation{School of Astronomy and Space Science, University of Chinese Academy of Sciences, Beijing 101408, China}
\email[]{xxx@nao.cas.cn}

\author[]{Xi Tan}
\affiliation{National Astronomical Observatories, Chinese Academy of Sciences, Beijing 100101, China}
\affiliation{School of Astronomy and Space Science, University of Chinese Academy of Sciences, Beijing 101408, China}
\email[]{xxx@nao.cas.cn}

\author[]{Yunao Xiao}
\affiliation{National Astronomical Observatories, Chinese Academy of Sciences, Beijing 100101, China}
\affiliation{School of Astronomy and Space Science, University of Chinese Academy of Sciences, Beijing 101408, China}
\email[]{xxx@nao.cas.cn}

\author[]{Jingyi Zhang}
\affiliation{National Astronomical Observatories, Chinese Academy of Sciences, Beijing 100101, China}
\email[]{xxx@nao.cas.cn}

\begin{abstract}
We present a comprehensive analysis of the physical properties of galaxies in the 
Dark Energy Spectroscopic Instrument (DESI) Data Release 1 (DR1) Bright Galaxy Survey (BGS), 
based on full spectral fitting of $\sim 6.7$ million galaxy spectra. 
Using a customized spectral fitting pipeline, we derive key physical parameters 
including stellar mass, stellar velocity dispersion, stellar population age, 
dust attenuation, and emission-line properties. 
To quantify the reliability and systematic uncertainties of our measurements, 
we construct a large set of mock spectra that closely reproduce the observed 
properties of DESI data, including realistic noise and spectral features. 
By comparing the recovered parameters with the known inputs, 
we assess the performance of the spectral fitting as a function of 
stellar continuum signal-to-noise ratio 
(S/N, defined as the ratio of the median continuum flux to 
its associated error)
and redshift. 
We find that stellar masses can be robustly recovered with negligible bias 
for spectra with $\mathrm{S/N} \gtrsim 5$, while low-S/N spectra 
($\mathrm{S/N} \lesssim 5$) show a mild systematic overestimation of 
$\sim 0.1$ dex and increased scatter. 
Similar trends are observed for stellar population parameters, 
while emission-line fluxes are recovered with high accuracy and minimal bias. 
We further validate our stellar mass estimates by comparison with independent 
measurements from photometric spectral energy distribution fitting, 
finding good overall consistency within the expected systematic uncertainties. 
The value-added catalog presented in this work enables a wide range of 
statistical studies of galaxy evolution with DESI, 
and provides a foundation for future analyses.
\end{abstract}

\section{Introduction} \label{sec:intro}

Galaxy spectra provide one of the most powerful tools for studying the physical 
properties of galaxies and their formation and evolution. The observed spectrum 
of a galaxy encodes rich information about its stellar populations, 
interstellar medium, and ionized gas, including stellar mass, 
star formation history, metallicity, dust attenuation, 
and emission-line properties. 
Extracting these physical quantities from galaxy spectra is therefore 
a cornerstone of modern extragalactic astronomy.

Over the past decades, a number of large spectroscopic surveys have revolutionized 
our understanding of galaxies by providing statistically significant samples 
with homogeneous data quality. Early surveys such as the Sloan Digital Sky Survey 
\citep[SDSS;][]{York_DG_2000} have enabled detailed studies of galaxy populations 
in the local universe, while subsequent surveys including GAMA \citep{Baldry_IK_2018},
VIPERS \citep{Scodeggio_M_2018}, 
and DESI have significantly expanded both the sample size and redshift coverage. 
In particular, the Dark Energy Spectroscopic Instrument \citep[DESI;][]{DESI_2016a,DESI_2016b,DESI_2022}
is conducting one of the largest spectroscopic surveys to date, 
aiming to obtain spectra for tens of millions of galaxies and quasars over a wide redshift range. 
The unprecedented scale of DESI provides a unique opportunity to study galaxy evolution 
with high statistical precision.

To fully exploit the scientific potential of such large datasets, 
robust methods are required to derive physical properties from galaxy spectra. 
A commonly adopted approach is full spectral fitting, in which the observed spectrum 
is modeled as a combination of simple stellar population (SSP) templates, 
together with prescriptions for dust attenuation and emission line components. 
A number of widely used spectral fitting codes have been developed for this purpose, 
including \texttt{pPXF} \citep{Cappellari_M_2004,Cappellari_M_2017},
\texttt{STARLIGHT} \citep{Cid-Fernandes_R_2005}, and more recent tools such as \texttt{Firefly} 
\citep{Wilkinson_DM_2015,Wilkinson_DM_2017}. 
These methods have been extensively applied to spectroscopic surveys 
(e.g., SDSS and Mapping nearby Galaxies at Apache Point Observatory \citep[MaNGA;][]{Bundy_K_2015}) 
to derive stellar population properties and emission line measurements.

Despite these advances, measuring galaxy physical properties from spectra remains challenging. 
One of the major difficulties is the well-known degeneracy between stellar population parameters 
(e.g., age and metallicity) and dust attenuation. 
In conventional spectral fitting, dust attenuation is typically modeled using an assumed parametric attenuation law, 
which can introduce systematic uncertainties and biases in the derived parameters. 
In recent years, alternative approaches have been proposed to mitigate these degeneracies, 
including methods that attempt to constrain dust attenuation in a more flexible or data driven way
\citep[e.g.,][]{Wilkinson_DM_2015,Li_N_2020}.
Another important challenge arises from the quality of the spectra. 
While surveys such as MaNGA provide high S/N integral field spectra for nearby galaxies, 
the BGS spectra typically have relatively low continuum S/N, 
especially for faint targets and at higher redshift, 
due to the bright-time observing conditions and short exposure times.
A substantial fraction of galaxy spectra have S/N values below $\sim 5$, 
which significantly limits the applicability of methods developed for high quality data. 
This necessitates the development of robust spectral fitting techniques 
that can reliably extract physical information from low S/N spectra.

The DESI Bright Galaxy Survey \citep[BGS;][]{Hahn_CH_2023}
provides a particularly valuable dataset for studying galaxy properties, 
with a large sample of galaxies spanning a wide range of stellar mass and star formation activity. 
Recent efforts have begun to construct value-added catalogs of physical properties 
for DESI galaxies using spectral energy distribution (SED) fitting and related techniques 
\citep[e.g.,][]{Siudek_M_2024,Zou_H_2024}. 
In addition, the DESI collaboration has developed the \texttt{FastSpecFit} pipeline \citep{Moustakas_J_2023}, 
which derives physical parameters through efficient modeling of galaxy spectra, 
and can also incorporate photometric information when available. 
These approaches have demonstrated the great potential of DESI data for statistical studies of galaxy populations.
However, several important issues remain. 
First, although current methods are able to extract basic physical parameters for large samples, 
the relatively low S/N of DESI spectra pose significant challenges for robust spectral fitting 
and may introduce non-negligible biases in the derived parameters. 
Second, a systematic assessment of these uncertainties under realistic observational conditions is still lacking, 
especially for methods based purely on optical spectral fitting. 
In this work, we aim to address these issues by performing a full spectral fitting analysis of DESI BGS galaxies, 
with a particular focus on quantifying the reliability of derived physical parameters using extensive mock spectra 
that closely match the properties of the data. 
This work serves as a first step toward a more comprehensive analysis of DESI spectra. 
In future work, we will combine spectra of galaxies with similar properties to construct high S/N stacked spectra, 
which will enable the application of more advanced techniques that do not rely on an assumed dust attenuation law, 
and thus allow us to further investigate the dust attenuation properties of galaxies.

In this work, we present a systematic analysis of the physical properties of galaxies 
in the DESI DR1 Bright Galaxy Survey based on full spectral fitting. 
Given the relatively low S/N of DESI spectra, we adopt a conventional approach 
in which the dust attenuation is described by a parameterized attenuation law during the fitting process, 
rather than attempting to measure it independently as in previous high-S/N studies. 
To quantify the reliability and potential biases of our measurements, 
we construct a large set of mock spectra that closely mimic the properties of the DESI data, 
and perform extensive tests of the spectral fitting procedure.

This paper is organized as follows. In Section~\ref{sec:data} we describe the DESI data and sample selection. 
In Section~\ref{sec:result} we show the main results on galaxy physical properties. 
In Section~\ref{sec:mock} we assess the robustness of our measurements using mock spectra. 
Finally, we discuss issues and summarize our conclusions in Section~\ref{sec:diss}.
Throughout this paper we assume a $\Lambda$ cold  dark matter cosmology model 
with $\Omega_m = 0.3$, $\Omega_{\Lambda} = 0.7$, and $H_0 = 70 \,km\,s^{-1}\,Mpc^{-1}$, 
and the stellar initial mass function  (IMF) of \cite{Chabrier_G_2003}.

\section{Data and Measurements} \label{sec:data}
\subsection{DESI DR1}

The Dark Energy Spectroscopic Instrument (DESI) is a new generation ground-based 
spectroscopic survey designed to probe the expansion history of the Universe 
and the growth of large-scale structure with unprecedented precision, 
and thereby constrain the nature of dark energy 
\citep{DESI_2013, DESI_2016a, DESI_2016b,DESI_2024a, DESI_2024b}.
DESI is expected to measure accurate redshifts for tens of millions of galaxies 
and quasars over a footprint of $\sim14{,}000~\mathrm{deg}^2$.

The instrument is mounted on the Mayall 4-m telescope at Kitt Peak National Observatory 
and features a $3.2^\circ$ field of view with 5000 robotically controlled fibers, 
enabling simultaneous observations of 5000 targets \citep{DESI_2022, Miller_TN_2024, Siber_JH_2023}. 
The fibers feed ten spectrographs, each with three arms (blue, red, and near-infrared), 
providing continuous wavelength coverage from $3600$ to 9800 \AA\
with a spectral resolution of $R\sim2000$--$5000$.
This resolution is sufficient to resolve key spectral features such as the [O\,II] doublet.
DESI targets are primarily selected from the DESI Legacy Imaging Surveys \citep{Zou_H_2017, Dey_A_2019}, 
which provide deep optical imaging in the $g$, $r$, and $z$ bands 
over $\sim20{,}000~\mathrm{deg}^2$, complemented by infrared data 
from the Wide-field Infrared Survey Explorer (WISE) mission. 
The imaging depth is significantly deeper than that of SDSS, 
enabling efficient target selection over a wide range of redshifts.

The DESI survey includes multiple target classes optimized for observations 
under different observing conditions. 
During dark time, the primary targets are luminous red galaxies (LRGs), 
emission-line galaxies (ELGs), and quasars (QSOs), 
spanning a wide redshift range up to $z\sim3.5$ 
\citep{Raichoor_A_2020, Ruiz-Macias_O_2020, Yeche_C_2020, Zhou_RP_2020,
Raichoor_A_2023, Chaussidon_E_2023, Zhou_RP_2023}. 
During bright time, DESI conducts the Bright Galaxy Survey (BGS), 
targeting low-redshift galaxies ($z\lesssim0.6$), 
as well as a Milky Way survey of stars \citep{Allende_PC_2020, Cooper_AP_2023, Hahn_CH_2023}.
DESI observations began with a survey validation (SV) phase prior 
to the start of the main survey in May 2021, 
followed by continuous data acquisition and processing 
through an automated pipeline \citep{Guy_J_2023, DESI_2024a}. 
The pipeline performs spectral extraction, wavelength and flux calibration, and redshift determination. 
Redshifts and classifications are obtained using the \texttt{Redrock} pipeline, 
which fits spectral templates of stars, galaxies, and quasars to the observed spectra
\citep{Brodzeller_A_2023}.

In this work, we use data from the first public data release of DESI 
\citep[DR1;][]{DESI_2025}, which provides a large sample of spectroscopically confirmed galaxies 
with well characterized data products. 
The DR1 dataset enables detailed statistical studies of galaxy properties based on optical spectroscopy.

\subsection{Bright Galaxy Survey Sample} \label{sec:data_bgs}

The Bright Galaxy Survey (BGS) is designed to obtain a highly complete sample 
of bright galaxies in the lower redshift universe ($z\lesssim0.6$), 
observed during bright observing conditions \citep{Hahn_CH_2023}.
The targets are primarily selected based on optical magnitude limits, 
resulting in a dense sample that spans a wide range of stellar mass,
star formation activity, and galaxy types.

\begin{figure*}
    \plotone{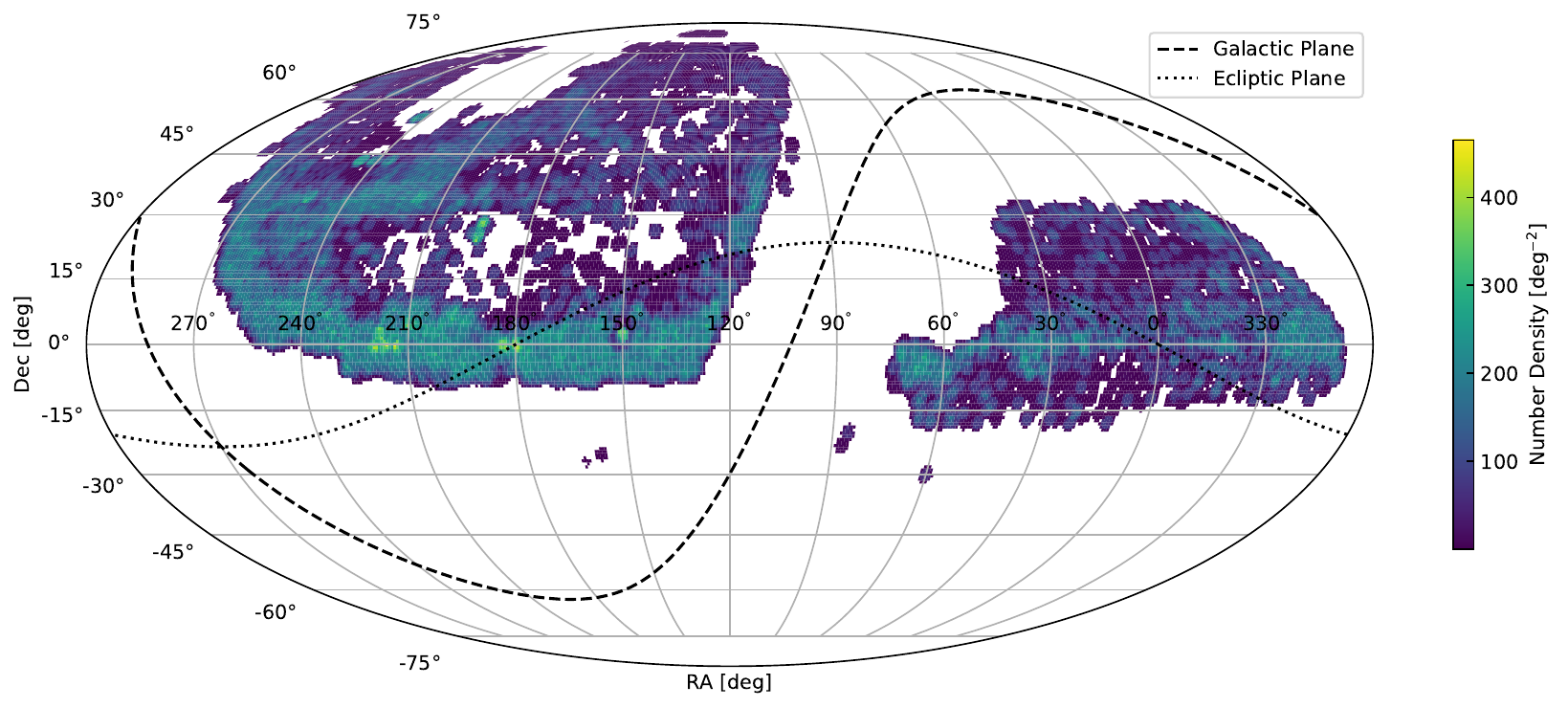}
    \caption{Sky distribution of the BGS sample in equatorial coordinates shown in a Mollweide projection. 
             The color scale indicates the galaxy number density in units of deg$^{-2}$. 
             The dashed and dotted curves denote the Galactic plane and the ecliptic plane, respectively.}
    \label{fig:ra_dec_dist}
\end{figure*}

In this work, we select galaxies from the BGS sample in DESI DR1 
based on the DESI targeting bitmasks. Specifically, BGS targets are identified 
using the \texttt{BGS\_ANY} flag in the \texttt{DESI\_TARGET} column for the main survey, 
as well as in the corresponding survey validation columns (\texttt{SV1}, \texttt{SV2}, and \texttt{SV3}). 
We further require reliable redshift measurements by applying the following quality cuts:
\texttt{ZWARN} = 0, \texttt{DELTACHI2} $>$ 25 \citep{Hahn_CH_2023}.
In addition, we restrict the sample to objects classified as galaxies based on their spectra.
After applying these criteria, we obtain a parent sample of 7,033,357 BGS spectra. 
Since DESI includes repeated observations of the same objects, 
we construct a sample of independent sources by selecting unique \texttt{TARGETID}, 
resulting in 6,673,756 galaxies. Among these, 6,662,812 galaxies have valid redshifts 
and successful spectral fitting.
Figure~\ref{fig:ra_dec_dist} shows the sky distribution of the BGS sample 
in equatorial coordinates using a Mollweide projection. The color map represents 
the number density of galaxies in units of deg$^{-2}$, highlighting the
non-uniform but contiguous coverage of the DESI footprint.

Due to the survey strategy and exposure time, 
the S/N of individual DESI spectra is typically modest. 
In particular, a large fraction of BGS spectra have S/N$\lesssim5$. 
In our sample, 3,817,100 galaxies have S/N$>5$, 
while the remaining sources are dominated by low S/N spectra. 
This characteristic of the data poses significant challenges 
for spectral fitting and motivates this work.

\subsection{Spectral Fitting Methodology}

To derive the physical properties of galaxies from the DESI BGS spectra,
we perform full spectral fitting using a customized pipeline developed in previous work
\citep{Li_N_2020, Li_N_2021, Li_N_2023}, with modifications tailored to the characteristics of DESI data.
The overall procedure consists of three main steps:
(i) modeling the stellar continuum and measuring stellar kinematics,
(ii) fitting and subtracting emission lines,
and (iii) deriving stellar population properties from the best-fit models.
Compared to our previous applications to high S/N integral field spectra,
we adopt a simplified treatment of dust attenuation appropriate for the relatively
low S/N of DESI spectra.

Before performing spectral fitting, we apply several preprocessing steps to the DESI spectra.
First, we correct for foreground Milky Way dust extinction using the $E(B-V)$ values 
provided in the DESI catalog, adopting the extinction curve of \citet{Cardelli_JA_1989}. 
Second, all spectra are shifted to the rest frame using the redshifts 
measured by the DESI pipeline. Given the high accuracy of DESI redshift 
measurements, we fix the systemic velocity to $v_* = 0$ during the fitting.
Third, we match the spectral resolution of the DESI data to that of the SSP models.
The BC03 templates have a spectral resolution of $\sim 3$ \AA, 
while the DESI spectra have a higher resolution corresponding to $\sim 1.8$ \AA\ on average.
We therefore convolve the DESI spectra with a Gaussian kernel to degrade 
their resolution to match the SSP models.
Finally, the spectra are rebinned onto a logarithmic wavelength grid 
in the rest frame, which facilitates the convolution with the velocity 
broadening kernel during fitting.

\subsection{Stellar Population Models}

We model the stellar component of galaxy spectra using simple stellar population (SSP)
models from \citet{Bruzual_Charlot_2003} (hereafter BC03).
These models are constructed based on the evolutionary tracks of
\citet{Bertelli_G_1994} and assume a \citet{Chabrier_G_2003} initial mass function (IMF).
The BC03 library contains 1326 SSPs spanning 221 ages from $t=0$ to $2\times10^{10}$ yr
and six metallicities in the range $Z=0.0001$ to $Z=0.05$.

To reduce computational cost while retaining the essential spectral information,
we apply principal component analysis (PCA) to the full SSP library,
following the method described in \citet{Li_C_2005}.
The first 9 eigenspectra are adopted as basis templates in the fitting.
These eigenspectra capture more than 99.9\% of the variance of the original SSP library,
and thus provide an efficient representation of the stellar population models.

\subsection{Stellar Continuum Fitting and Kinematics}\label{sec:continuum_fit}

We first fit the observed spectrum with a linear combination of the PCA eigenspectra
to model the stellar continuum and measure stellar kinematics.
The model spectrum is given by
\begin{equation}
F(\lambda) = \left(\sum_{j=1}^{9} x_j f^j_{\rm PCA}(\lambda) A(\lambda)\right)
\otimes G(v_*, \sigma_*),
\end{equation}
where $f^j_{\rm PCA}(\lambda)$ is the $j$-th eigenspectrum,
$x_j$ are the coefficients to be determined,
$G(v_*, \sigma_*)$ is a Gaussian kernel representing the line-of-sight velocity distribution,
and $A(\lambda)$ describes dust attenuation.
Given that the spectra have been shifted to the rest frame using accurate DESI redshifts,
we fix the systemic velocity to $v_* = 0$ and only fit for the velocity dispersion $\sigma_*$.

During the fitting, strong emission lines are masked iteratively
to avoid contamination of the stellar continuum fit.
Given the relatively low S/N of DESI spectra, we adopt a conventional approach 
in which dust attenuation is modeled using a parametric attenuation law.
Specifically, we assume the \citet{Calzetti_D_2000} attenuation curve,
with the color excess $E(B-V)$ as a free parameter in the fitting.
This differs from our previous work on MaNGA data,
where the attenuation curve was measured in a model-independent manner.
Such an approach is not robust for spectra with S/N $\lesssim 5$,
where noise strongly affects the separation of large- and small-scale spectral features,
as demonstrated in previous studies \citep{Li_N_2020},
and therefore is not adopted here.
Therefore, in this work we adopt a conventional parametric attenuation law
during spectral fitting, and focus on assessing the reliability
of the derived physical parameters using mock spectra,
as described in Section~\ref{sec:result}.

\subsection{Emission Line Measurements}

After subtracting the best-fit stellar continuum, we fit the emission lines in the residual spectrum.
We include a set of commonly used optical emission lines,
such as H$\alpha$, H$\beta$, [O\,III], [N\,II], [S\,II], and [O\,II].

Each emission line is modeled with a Gaussian profile.
For the [O\,II] $\lambda\lambda3726,3729$ doublet, the two components are well resolved in DESI spectra,
and are therefore fitted independently without fixing their flux ratio.
For the [N\,II] $\lambda\lambda6548,6584$ doublet, we fix the flux ratio to the theoretical value of 3.
Similarly, for the [O\,III] $\lambda\lambda4959,5007$ doublet, the flux ratio is fixed to 3.
For the [S\,II] $\lambda\lambda6717,6731$ doublet, we allow the flux ratio to vary freely, 
as it is sensitive to electron density.
The fitting yields the flux, equivalent width,
velocity, and velocity dispersion of each emission line.

\subsection{Stellar Population Properties}

Using the best-fit stellar continuum model,
we derive stellar population properties through full spectral fitting.
We construct a reduced SSP library consisting of 150 templates,
selected from the BC03 models to cover 25 logarithmically spaced ages
for each of the six metallicities. The observed spectrum is then fitted as
\begin{equation}
F(\lambda) = \sum_{j=1}^{150} x_j f^j_{\rm SSP}(\lambda),
\end{equation}
where $x_j$ represents the contribution of each SSP.

From the best-fit coefficients, we derive physical properties
including stellar mass, light-weighted and mass-weighted stellar ages, and stellar metallicities.
These quantities are computed following standard definitions based on the SSP weights.

\section{Results on observed spectra} \label{sec:result}

In this section, we present the derived physical properties of the DESI BGS sample 
based on our full spectral fitting pipeline. 
We summarize the global distributions of the fundamental galaxy parameters, 
assess the fitting quality across different redshift ranges, 
and validate our stellar mass measurements by comparing them with existing value-added catalogs.

\subsection{Sample Statistics and Fitting Quality}

\begin{figure}
    \epsscale{1.15}
    \plotone{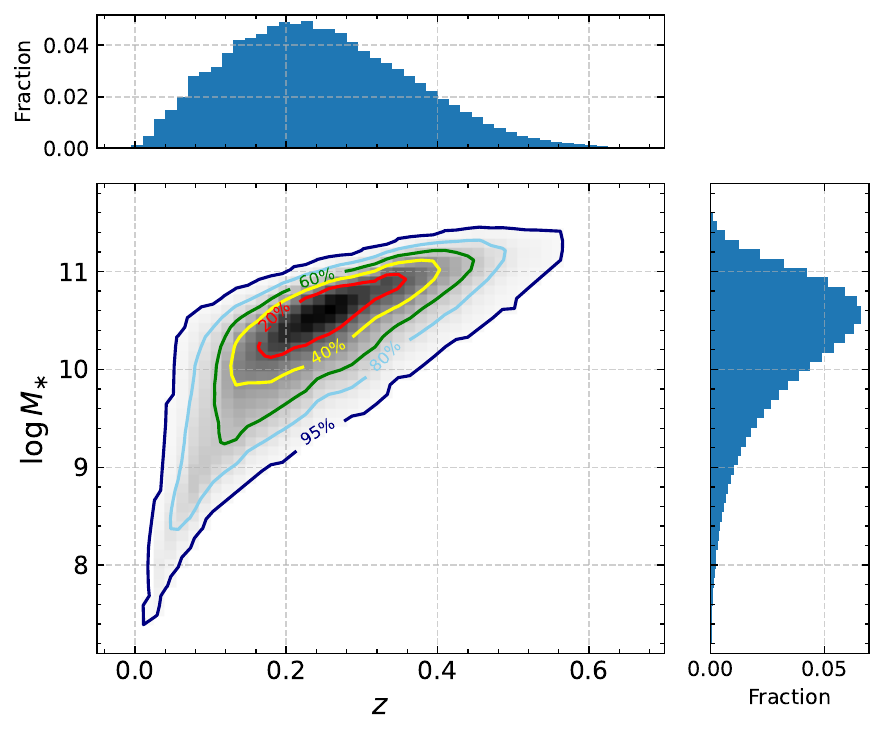}
    \caption{Distribution of galaxies in the stellar mass--redshift ($M_{\ast}$--$z$) plane 
            for the full sample of BGS galaxies with successful spectral fitting. 
            The color scale represents the number density of galaxies in each bin. 
            The top and right panels show the distributions of redshift and stellar mass, respectively. 
            Overplotted contours indicate regions enclosing 95\%, 80\%, 60\%, 40\%, and 20\% 
            of the total galaxy population, as derived from the cumulative two-dimensional density distribution. 
            A clear trend is seen in which low-mass galaxies dominate at low redshift, 
            while the high-redshift regime is increasingly populated by massive galaxies, 
            reflecting the flux-limited selection of the BGS sample. }
    \label{fig:z_M_hist}
\end{figure}

After applying the target selection and quality control criteria described in Section \ref{sec:data_bgs}, 
our final robust sample comprises 6,662,812 unique galaxies with valid redshifts 
and successful full spectral fits. 
Figure~\ref{fig:z_M_hist} shows the distribution of galaxies in the stellar mass--redshift ($M_{\ast}$--$z$) plane. 
The sample spans a wide range in stellar mass, from $\log M_{\ast} \sim 7$ to $\sim 12$, 
and covers redshifts up to $z \sim 0.6$. 
At low redshift, the sample includes a large population of low-mass galaxies, 
while at higher redshift it becomes increasingly dominated by massive systems, 
reflecting the flux-limited nature of the BGS selection \citep{Hahn_CH_2023}.

\begin{figure*}
    \epsscale{1.15}
    \plotone{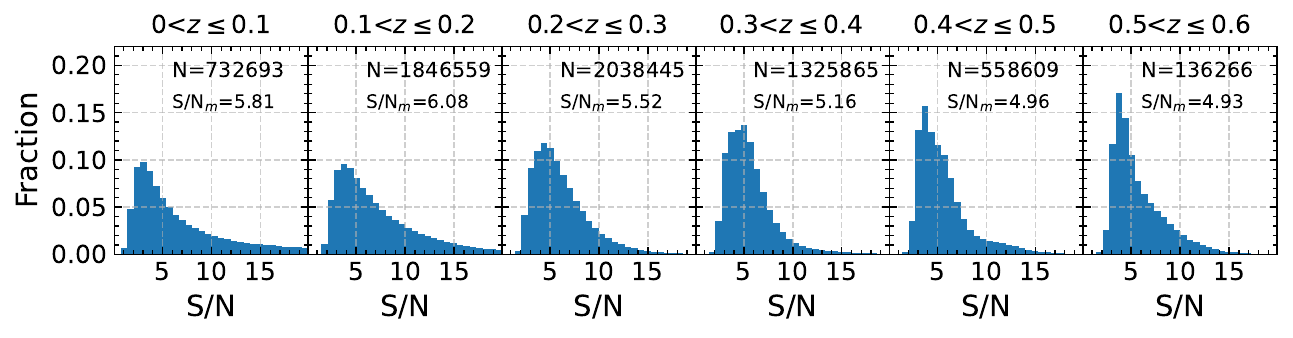}
    \plotone{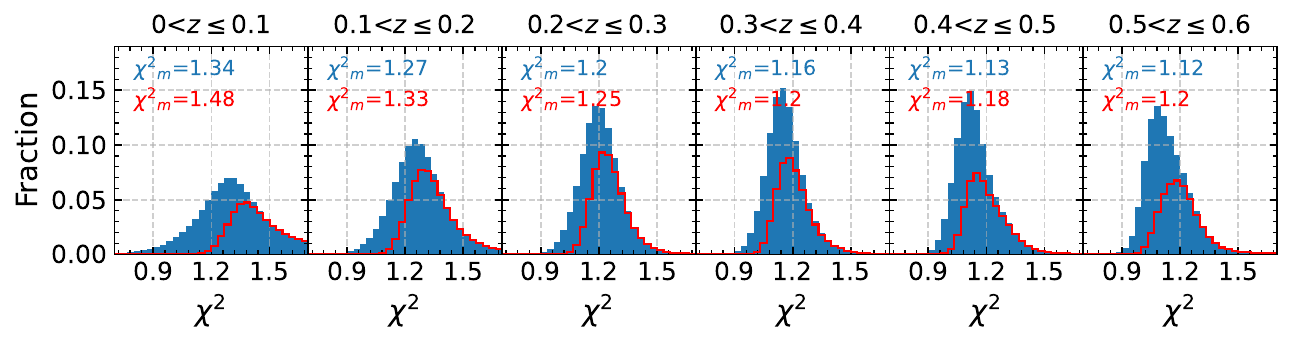}
    \caption{Distributions of spectral S/N (top panels) 
        and reduced $\chi^2$ of the continuum fitting (bottom panels) 
        in six redshift bins spanning $0 < z \leq 0.6$. 
        Each panel shows the normalized histogram of galaxies in the corresponding redshift interval.
        In the top panels, the total number of galaxies ($N$) and the median S/N (${\rm S/N}_m$) 
        are indicated for each redshift bin. 
        The S/N distributions show a mild decrease with increasing redshift, 
        while remaining relatively stable around $\sim 5$--6 across the full sample.
        In the bottom panels, the red histograms show the high-quality subsample with S/N $> 5$,
        and the median reduced $\chi^2$ values for the full sample 
        (blue) and for the subsample with S/N $> 5$ (red) are indicated. 
        The reduced $\chi^2$ is calculated based solely on the continuum windows,
        with emission-line regions masked during the fit.
    }
    \label{fig:hsit_snr_vd}
\end{figure*}

As discussed in Section~\ref{sec:intro}, a major challenge of the DESI dataset 
is the relatively modest S/N of individual spectra. 
In our sample, 3,817,100 galaxies have S/N $> 5$ 
and 6,022,027 galaxies have S/N $> 3$,
while a non-negligible fraction remains at lower S/N.
To quantify the data quality and assess the robustness of the spectral fitting, 
Figure~\ref{fig:hsit_snr_vd} shows the distributions of the S/N and the reduced $\chi^2$ 
of the continuum fitting in six redshift bins spanning $0 < z \leq 0.6$.
Note that this reduced $\chi^2$ is computed only for the continuum windows, 
after masking the strong emission lines as described in Section~\ref{sec:continuum_fit}.

The S/N distributions (top panels of Figure~\ref{fig:hsit_snr_vd}) 
show a mild but systematic decline with increasing redshift. 
The median S/N decreases from $\sim 5.8$ at $z \leq 0.1$ to $\sim 4.9$ at $z > 0.5$, 
reflecting the increasing distance and the approximately fixed exposure time of the survey. 
Despite this trend, the S/N remains relatively stable across the full redshift range, 
with typical values around $\sim 5$--6. 
The number of galaxies peaks at intermediate redshift ($z \sim 0.1$--0.3), 
consistent with the selection function of the BGS sample.

The distributions of reduced $\chi^2$ (bottom panels of Figure~\ref{fig:hsit_snr_vd}) 
show a clear peak near unity in all redshift bins, 
indicating that the spectral fitting model provides a good 
overall description of the observed spectra. 
The median $\chi^2$ decreases slightly from $\sim 1.34$ at low redshift 
to $\sim 1.12$ at $z > 0.5$. 
This trend is consistent with the decreasing S/N, as lower-quality spectra 
naturally yield $\chi^2$ values closer to unity.
We further compare the $\chi^2$ distributions for the full sample and for a 
high-quality subsample with S/N $> 5$. 
For the high-S/N subsample, the median $\chi^2$ values are systematically higher, 
ranging from $\sim 1.48$ at low redshift to $\sim 1.20$ at high redshift. 
This behavior reflects the fact that higher-S/N spectra place stronger constraints 
on the models and are more sensitive to small mismatches between the data and templates. 
Overall, the absence of significant systematic deviations from unity 
demonstrates the robustness of our fitting procedure across a wide range of data quality.

\subsection{Distributions of Physical Properties}

\begin{figure*}
    \epsscale{1.05}
    \plotone{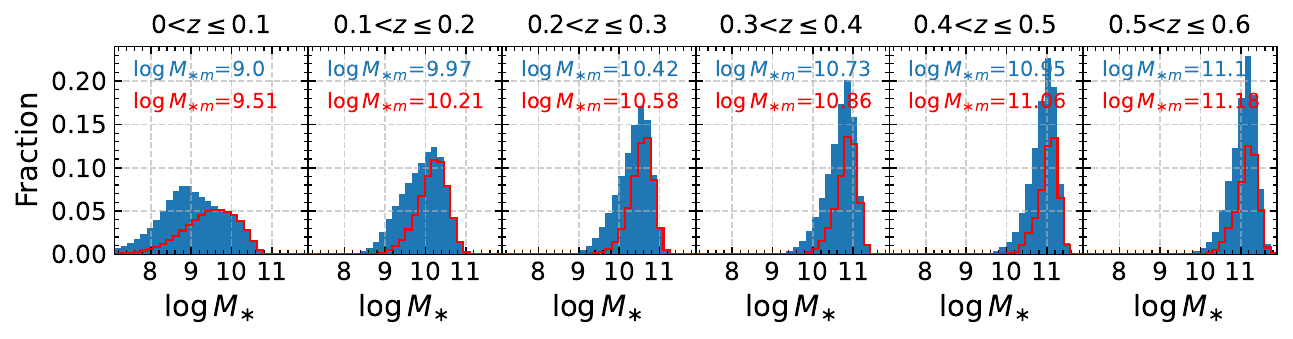}
    \plotone{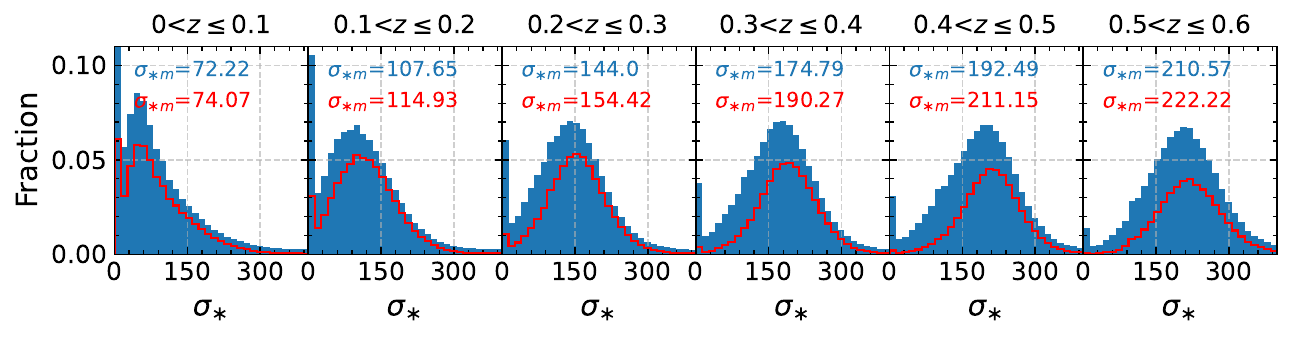}
    \plotone{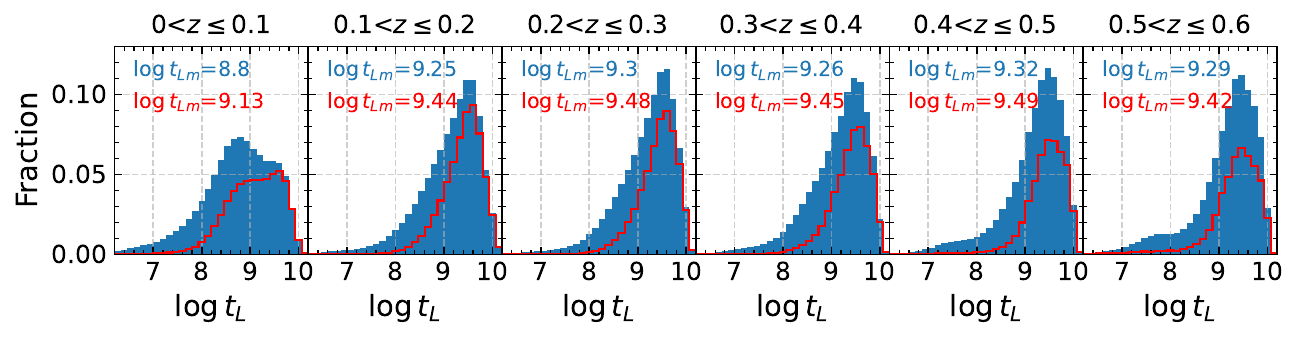}
    \plotone{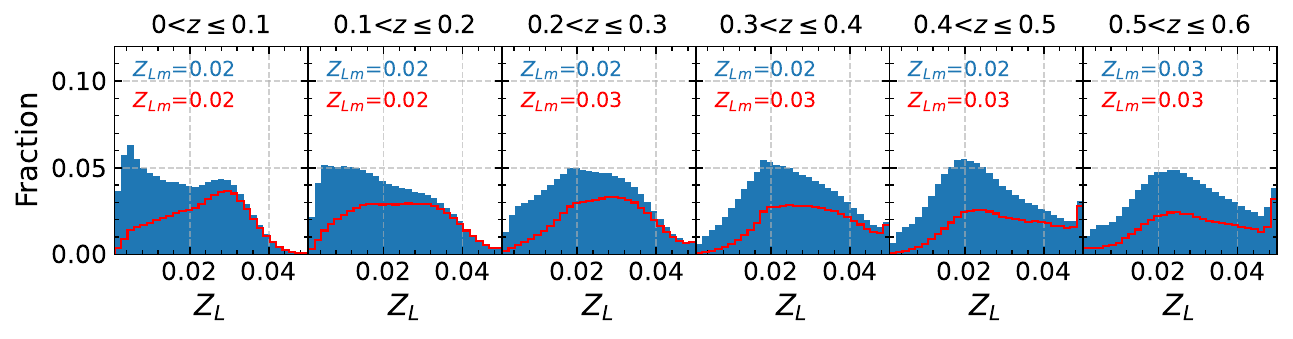}
    \plotone{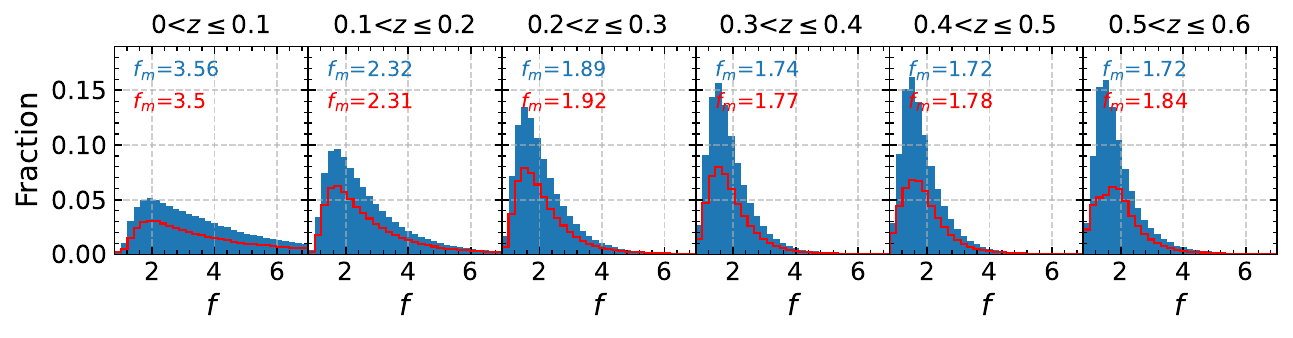}
    \caption{Distributions of $M_{\ast}$, $\sigma_{\ast}$, $t_L$, $Z_L$, and $f$
        in six redshift bins spanning $0 < z \leq 0.6$. 
        Each panel shows the normalized histogram of galaxies in the corresponding redshift interval.
        For each parameter, the median values for the full sample (blue) and for the 
        high-quality subsample with S/N $> 5$ (red) are indicated in each panel. 
        The high-S/N subsample is also shown as red histograms for comparison.
    }
    \label{fig:hsit_m_t_f}
\end{figure*}

Using the methodology outlined in Section~\ref{sec:data}, 
we derive a set of key physical parameters, including stellar mass ($M_{\ast}$), 
stellar velocity dispersion ($\sigma_{\ast}$), light-weighted stellar age ($t_L$), 
light-weighted stellar metallicity ($Z_L$),
and a flux calibration scaling factor ($f$) (see the definition in Section~\ref{subsec:aper_corr}). 
Figure~\ref{fig:hsit_m_t_f} presents the normalized distributions of these quantities 
in six redshift bins spanning $0 < z \leq 0.6$.

The stellar mass distributions show a strong systematic shift toward higher values 
with increasing redshift. The median stellar mass increases from 
$\log M_{\ast} \sim 9.0$ at $z \leq 0.1$ to $\sim 11.1$ at $z > 0.5$. 
For the high-quality subsample with S/N $> 5$, the median masses are systematically higher, 
ranging from $\sim 9.5$ to $\sim 11.2$. 
This trend is consistent with the flux-limited selection of the BGS sample, 
as also illustrated in Figure~\ref{fig:z_M_hist}.

The distributions of stellar velocity dispersion exhibit a similar trend, 
with median values increasing from $\sim 70$ km s$^{-1}$ at low redshift 
to $\sim 210$ km s$^{-1}$ at $z > 0.5$, reflecting the increasing dominance of massive galaxies. 
The high-S/N subsample shows systematically larger velocity dispersions. 
This is primarily a selection effect, as galaxies with higher S/N are typically 
more luminous and massive, and therefore tend to have intrinsically larger 
velocity dispersions. A similar behavior is seen for stellar mass and 
light-weighted age, reflecting the underlying correlation between galaxy 
luminosity, mass, and stellar population properties.
We note that a non-negligible fraction of spectra have $\sigma_{\ast} \approx 0$, 
which corresponds to cases where the velocity dispersion cannot be robustly constrained, 
typically due to low S/N or weak absorption features. 
These measurements are included in the distributions but should be interpreted with caution.

The light-weighted stellar age distributions are broad across all redshift bins, 
with median values around $\log t_L \sim 9.3$. 
The high-S/N subsample yields slightly older ages, 
with medians of $\log t_L \sim 9.4$--9.5, 
which is likely related to the underlying correlation between S/N, 
galaxy luminosity, and stellar population properties.
Overall, the weak evolution of $t_L$ reflects the mixture of star-forming 
and quiescent galaxies in the BGS sample.

The flux calibration scaling factor shows a decreasing trend with redshift, 
with median values declining from $\sim 3.6$ at low redshift to $\sim 1.7$ at higher redshift. 
This behavior likely reflects aperture and calibration differences between 
the spectroscopic and photometric measurements, which become less significant 
for more distant galaxies.

For clarity, only a subset of the derived parameters is shown here. 
The full set of measured physical properties is described in Appendix~\ref{app:data_model}.

\subsection{Aperture Corrections and Stellar Mass Consistency} \label{subsec:aper_corr}

Since DESI is a fiber-fed spectrograph, the physical properties derived directly 
from the spectra represent only the light enclosed within the fiber aperture 
(1.5 arcsec in diameter) \citep{DESI_2022, Miller_TN_2024, Siber_JH_2023}. 
To compute the total stellar mass of the galaxies, 
we calculate an aperture correction factor, $f$, defined as the ratio of 
the total photometric flux to the fiber flux. 
Using the DESI Legacy Imaging Surveys photometry \citep{Zou_H_2017, Dey_A_2019}, 
we estimate this factor by taking the average of the flux ratios in the $g$ and $r$ bands:
\begin{equation}
    f = \frac{1}{2} \left( \frac{F_{g,\mathrm{total}}}{F_{g,\mathrm{fiber}}} + \frac{F_{r,\mathrm{total}}}{F_{r,\mathrm{fiber}}} \right).
\end{equation}
The distribution of this calibration scale factor $f$ is shown in the bottom panel of 
Figure \ref{fig:hsit_m_t_f}. 
We can correct our fiber-based stellar masses to total stellar masses by adding $\log f$.
This aperture correction is based on the assumption that the stellar mass-to-light ratio
outside the fiber is similar to that measured within the fiber aperture,
that is, no significant color or stellar population gradients are considered.
In reality, galaxies often exhibit radial variations in stellar age, metallicity,
and dust attenuation, which can introduce systematic uncertainties into the total stellar mass estimates.
Nevertheless, given the relatively low S/N of most DESI spectra and the large statistical sample used in this work,
the simple flux-scaling correction provides a practical first-order approximation 
for converting fiber-based measurements to global galaxy properties.

\begin{figure*}
    \epsscale{1.15}
    \plotone{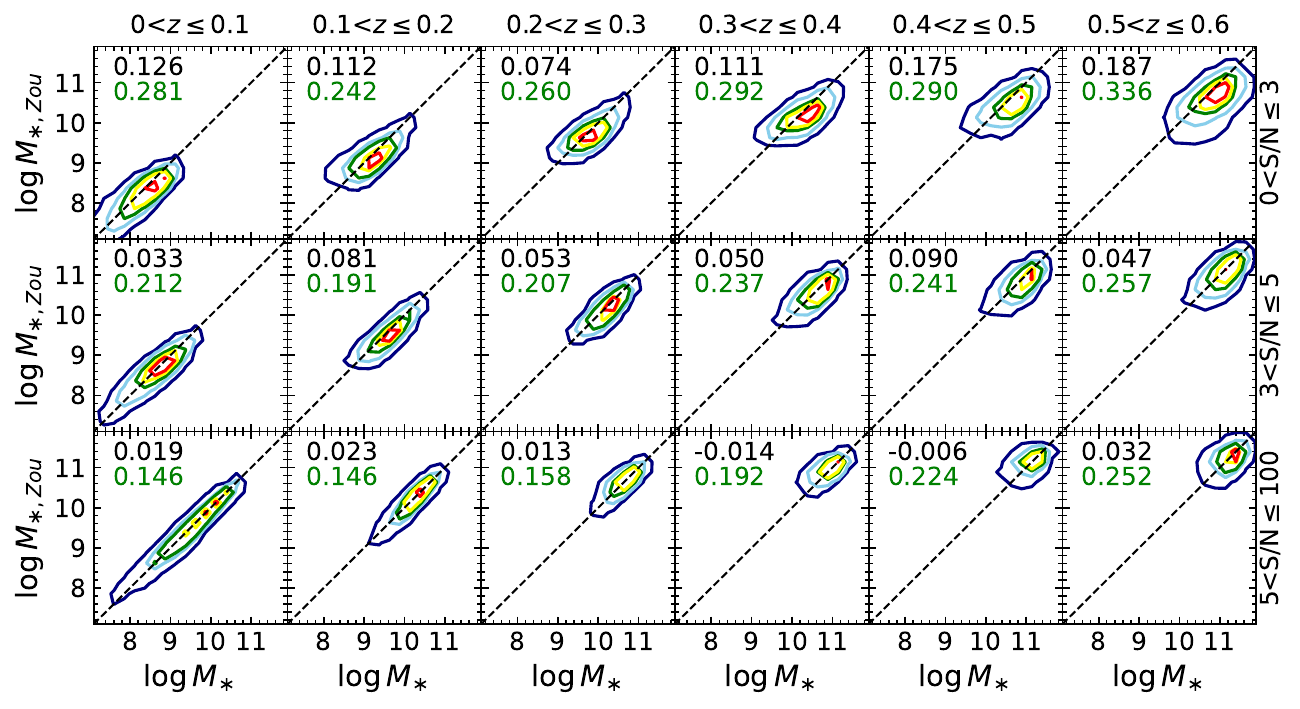}
    \caption{
        Comparison of stellar masses derived in this work with those from 
        the official DESI DR1 stellar mass VAC (based on \texttt{CIGALE} 
        photometric SED fitting; methodology described in \citet{Zou_H_2024}).
        Panels are divided into six redshift bins from $z=0$ to $z=0.6$ (left to right), 
        and three S/N regimes: S/N $\leq 3$ (top), $3 <$ S/N $\leq 5$ (middle) and S/N $> 5$ (bottom). 
        The horizontal axis shows $\log M_{\ast}$ derived in this work, 
        while the vertical axis shows $\log M_{\ast}$ from the VAC. 
        The dashed line indicates the one-to-one relation.
        Overplotted contours indicate regions enclosing 95\%, 80\%, 60\%, 40\%, and 20\% 
        of the total galaxy population, as derived from the cumulative two-dimensional density distribution. 
        In each panel, the median offset ($\Delta \log M_{\ast}$) (black) and standard deviation (green) 
        are indicated. 
    }
    \label{fig:M_zou}
\end{figure*}

To validate our measurements, we compare our derived total stellar masses 
with independent estimates from the literature. 
Specifically, we cross-match our sample with the official DESI DR1 
Value-Added Catalog (VAC) of Stellar Mass and Emission Line Catalog 
\footnote{https://data.desi.lbl.gov/doc/releases/dr1/vac/stellar-mass-emline/},
which provides photometric SED-fitting results derived using 
\texttt{CIGALE} \citep{Boquien_M_2019} for the full DESI DR1 sample. 
This VAC is constructed following the methodology described in \citet{Zou_H_2024}. 
Figure~\ref{fig:M_zou} shows the comparison between our stellar masses and 
those from this VAC, divided into six redshift bins and three S/N regimes.
Overall, the two measurements show good consistency over a wide mass range, 
with most galaxies distributed close to the one-to-one relation.
A clear dependence on S/N is observed. 
For the high-S/N subsample (S/N $> 5$), the median offset between the two measurements 
is small, typically within $\sim 0.02$ dex across all redshift bins, 
with a scatter of $\sim 0.15$--0.25 dex. 
In contrast, for the low-S/N subsample (S/N $\leq 3$), the offsets are slightly larger, 
with median differences of $\sim 0.11$ dex and increased scatter 
of $\sim 0.24$--0.34 dex.
These trends indicate that the agreement between spectroscopic and photometric 
stellar mass estimates depends on data quality, with larger scatter and small 
systematic offsets emerging in the low-S/N regime. 
Possible sources of these differences include uncertainties in spectral fitting 
at low S/N, as well as intrinsic differences between spectroscopic and photometric 
methods, such as the treatment of star formation histories, dust attenuation
and the aperture correction.

Overall, the level of agreement is comparable to that reported in previous studies, 
and supports the consistency of our stellar mass estimates within the expected 
uncertainties.
We note that systematic differences at the level of $\sim0.1$--$0.3$ dex 
are commonly found when comparing stellar mass estimates derived from different methods, 
even when based on similar datasets. 
For example, \citet{Siudek_M_2024} performed a comprehensive comparison 
between multiple value-added catalogs for DESI galaxies, 
including both spectral fitting based and photometric SED fitting approaches. 
They found median offsets ranging from $\Delta \sim -0.39$ to $+0.06$ dex 
and scatter of $\sim0.13$--$0.58$ dex across different catalogs, 
depending on the adopted models, star formation histories, and dust attenuation prescriptions. 
In particular, the comparison with \citet{Zou_H_2024} yields 
$\Delta \sim -0.10$ dex and scatter $\sim0.17$, 
while comparisons with other commonly used catalogs such as 
SDSS MPA--JHU \citep{Kauffmann_G_2003,Brinchmann_J_2004}
and GALEX-SDSS-WISE Legacy Catalog \citep[GSWLC;][]{Salim_S_2016,Salim_S_2018}
also show offsets at the level of $\sim0.1$--$0.2$ dex.
These variations arise from a combination of factors, including differences 
in stellar population synthesis models, assumed star formation histories, 
dust attenuation laws, and the use of spectroscopy versus broadband photometry. 
Given these known systematics, the level of agreement shown in Figure~\ref{fig:M_zou}, 
with median offsets typically $\lesssim0.1$ dex and comparable scatter, 
is fully consistent with expectations. 
This further supports the robustness of our stellar mass estimates 
derived from full spectral fitting of DESI spectra.

\subsection{BPT Diagram and Redshift Evolution}

\begin{figure*}
    \epsscale{1.15}
    \plotone{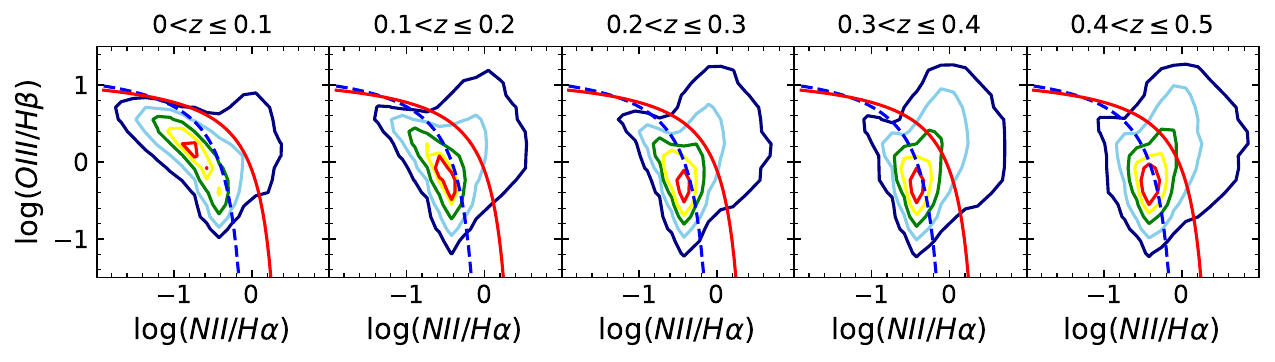}
    \caption{BPT diagnostic diagram ([O\,III] $\lambda5007$/H$\beta$ versus 
        [N\,II] $\lambda6583$/H$\alpha$) for galaxies in different redshift bins. 
        From left to right, the panels show galaxies in the redshift ranges 
        $z=0$--0.1, 0.1--0.2, 0.2--0.3, 0.3--0.4, and 0.4--0.5. 
        Only galaxies with reliable measurements of all four emission lines are included. 
        The blue dashed curve denotes the empirical demarcation between star-forming 
        and composite galaxies from \citet{Kauffmann_G_2003b}, 
        while the red solid curve shows the theoretical maximum starburst line 
        from \citet{Kewley_LJ_2001}. 
        Galaxies located below the blue curve are classified as star-forming systems, 
        those between the two curves as composite, and those above the red curve 
        as AGN-dominated systems. 
        At $z \gtrsim 0.5$, the H$\alpha$ and [N\,II] lines shift beyond the DESI 
        wavelength coverage, and are therefore not included in this analysis.}
    \label{fig:bpt}
\end{figure*}

Figure~\ref{fig:bpt} shows the distribution of galaxies 
on the Baldwin--Phillips--Terlevich diagram \citep[BPT;][]{Baldwin_JA_1981}
in different redshift bins.
At low redshift ($z\leq0.1$), the majority of galaxies lie below the 
\citet{Kauffmann_G_2003b} demarcation line, indicating that the sample 
is dominated by star-forming galaxies. 
These systems are characterized by relatively high 
[O\,III]/H$\beta$ and low [N\,II]/H$\alpha$ ratios, 
consistent with low-metallicity, high-ionization H\,II regions.
With increasing redshift, the distribution of galaxies gradually shifts 
toward lower [O\,III]/H$\beta$ and higher [N\,II]/H$\alpha$ ratios, 
moving along the star-forming sequence toward the composite and AGN regions. 
At $z \gtrsim 0.3$, a significant fraction of galaxies populate the region 
above the \citet{Kewley_LJ_2001} line, suggesting an increasing contribution 
from AGN or composite systems.

This trend can be understood as a combination of several effects. 
First, due to the flux-limited nature of the BGS sample, galaxies observed 
at higher redshift are preferentially more massive (see Figure~\ref{fig:z_M_hist}). 
More massive galaxies tend to have higher gas-phase metallicities and lower 
ionization parameters, which naturally lead to higher [N\,II]/H$\alpha$ 
and lower [O\,III]/H$\beta$ ratios \citep[e.g.,][]{Tremonti_CA_2004, Kewley_LJ_2013}. 
Second, massive galaxies are more likely to host active galactic nuclei, 
which contributes to the increased fraction of objects in the AGN region 
of the diagram. 
In addition, observational effects may also play a role. 
At higher redshift, the typical S/N of the spectra decreases, 
which can bias emission-line measurements and reduce the detectability 
of weak star-forming features. 
Furthermore, aperture effects become more significant, 
as the fixed fiber size samples a larger physical region of galaxies, 
potentially mixing nuclear and disk emission. 

Overall, the observed evolution in the BPT distribution is consistent with 
the combined effects of galaxy population changes and observational selection, 
rather than indicating an intrinsic disappearance of star-forming galaxies 
at higher redshift.
It is important to emphasize that the BPT diagrams shown here are intended 
to illustrate the distribution of the BGS sample in different redshift bins, 
rather than to trace the intrinsic redshift evolution of emission-line properties. 
As discussed above, the observed trends are significantly affected by 
selection effects inherent to the flux-limited survey, as well as 
observational factors such as varying S/N and aperture effects. 
A proper assessment of the redshift evolution of galaxy ionization properties 
would require carefully controlled samples and forward modeling of these effects, 
which is beyond the scope of this work.

\section{Tests on mock spectra} \label{sec:mock}

\subsection{Mock Spectra Construction and Validation} \label{sec:mock_spectra}

To assess the reliability and potential biases of our spectral fitting pipeline, 
we construct a large set of mock spectra that closely resemble the DESI BGS data. 
The mock spectra are generated based on the best-fit models of real galaxies, 
allowing us to perform a controlled forward-modeling test of the fitting procedure.

We begin by randomly selecting a subsample of 100,000 galaxies from the full BGS sample. 
The selection is performed from the set of unique sources with valid redshifts and 
successful spectral fitting (see Section~\ref{sec:data_bgs}), ensuring that the mock sample 
covers the same parameter space as the observed data. 
For each selected galaxy, we generate a mock spectrum using its best-fit stellar population 
and emission-line model. The stellar continuum is constructed as a linear combination of 
SSP templates weighted by the fitted coefficients, and is then convolved with the 
line-of-sight velocity distribution characterized by the fitted velocity dispersion. 
Dust attenuation is applied using the same parametric attenuation law adopted in the fitting.

\begin{figure*}
    \epsscale{1.15}
    \plotone{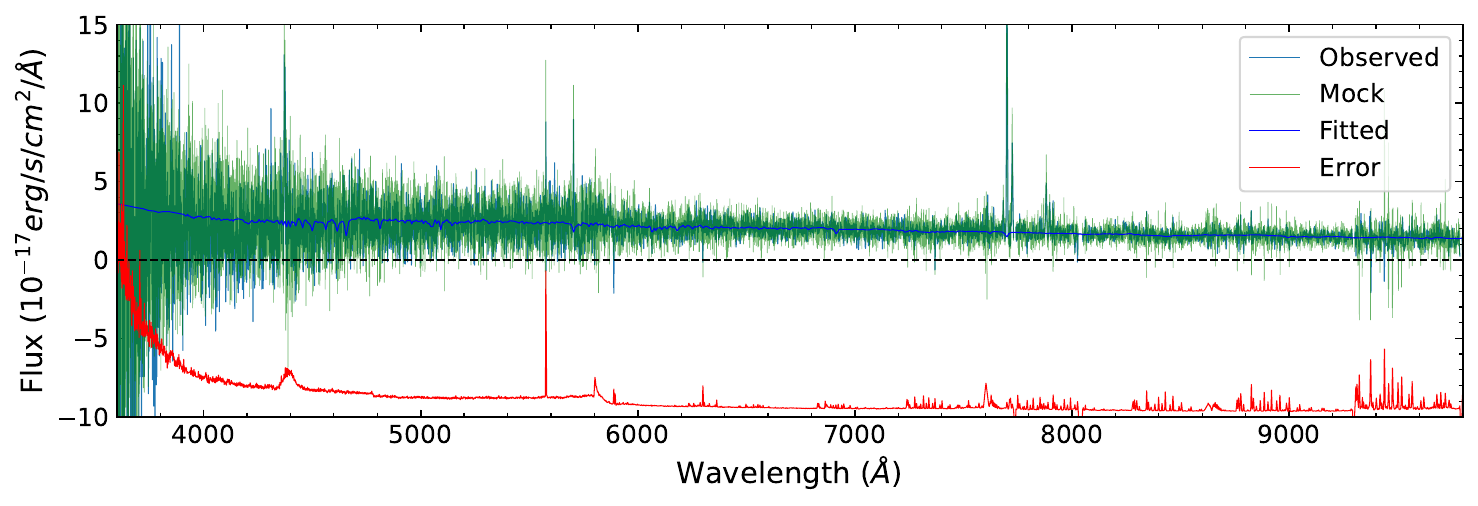}
    \caption{Example of a mock spectrum constructed from a DESI BGS galaxy. 
        The light blue curve shows the observed spectrum (corrected for Galactic extinction), 
        while the blue curve represents its best-fit model from spectral fitting. 
        The mock spectrum (green) is generated by adding emission-line components 
        and Gaussian noise, based on the observational uncertainties, to the best-fit model. 
        Because the mock spectrum closely follows the observed spectrum, 
        the latter is largely overplotted by the green curve. 
        The red curve shows the flux uncertainty, shifted downward by 10 units for clarity. 
        }
    \label{fig:example_mock}
\end{figure*}

To ensure that the mock spectra realistically reproduce the noise properties of DESI data, 
we extract the wavelength-dependent flux uncertainties from the corresponding observed spectra 
and use them to generate Gaussian noise. The noise is added to the model spectrum after 
rebinning to the observed wavelength grid. Emission lines are included by adding Gaussian 
profiles with fluxes and widths taken from the fitting results. 
In this way, each mock spectrum consists of three well-defined components: 
(i) a stellar continuum described by SSP models, 
(ii) nebular emission lines, and 
(iii) Gaussian noise consistent with the observational uncertainties. 
This construction ensures that the input true physical parameters are known, 
while preserving the realistic observational conditions of DESI spectra.
An example of the mock construction is shown in Figure~\ref{fig:example_mock}, 
where we compare the observed spectrum, the best-fit model, and the resulting mock spectrum. 
The close agreement between the observed and mock spectra demonstrates that our procedure 
successfully reproduces both the spectral features and noise characteristics of the data.

The resulting mock spectra are then processed using the same spectral fitting pipeline 
as applied to the observed data. By comparing the recovered parameters with the input values, 
we can directly quantify the accuracy and biases of our measurements as a function of 
spectral quality and galaxy properties.

\subsection{Stellar Mass Recovery from Mock Spectra}

Using the mock spectra constructed in Section~\ref{sec:mock_spectra}, 
we directly compare the recovered stellar masses from spectral fitting 
with the input values used to generate the mocks. 
This provides a quantitative assessment of both systematic biases 
and statistical uncertainties in our measurements. 
In this subsection, we focus on stellar mass as a representative example.

\begin{figure*}
    \epsscale{1.15}
    \plotone{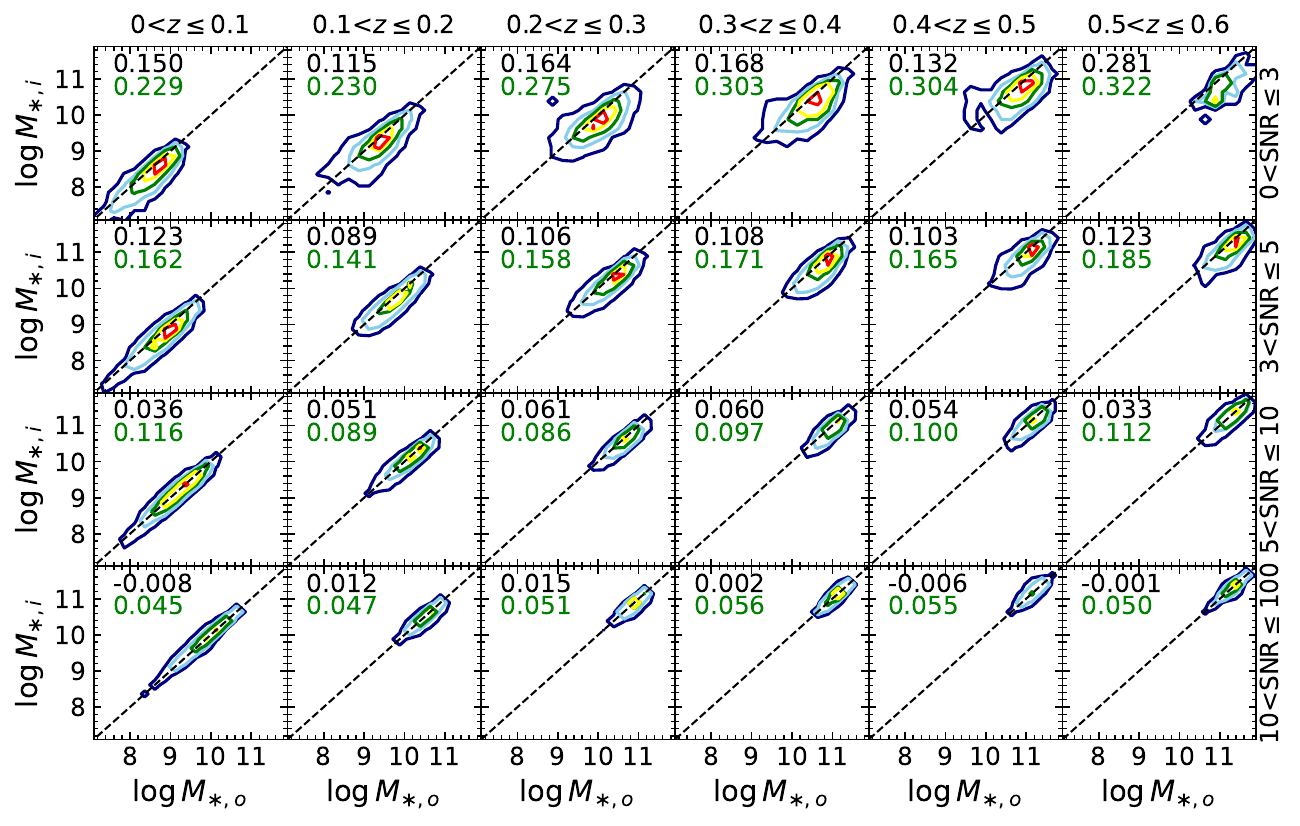}
    \caption{Comparison between the recovered stellar mass ($\log M_{\ast, o}$) 
        and the input stellar mass ($\log M_{\ast, i}$) for the mock spectra. 
        From left to right, the panels show six redshift bins from $z=0$ to 0.6. 
        From top to bottom, the rows correspond to four S/N ranges: 
        S/N $\leq 3$, $3 < \mathrm{S/N} \leq 5$, 
        $5 < \mathrm{S/N} \leq 10$, and S/N $> 10$. 
        Each panel shows the density distribution of galaxies, 
        with the dashed line indicating the one-to-one relation. 
        The median offset $\Delta \log M_* = \log M_{\ast, o} - \log M_{\ast, i}$ (black)
        and the standard deviation (green) are labeled in each panel. 
        A clear dependence on S/N is observed: low-S/N spectra show a systematic 
        overestimation of stellar mass and larger scatter, while high-S/N spectra 
        yield nearly unbiased measurements with significantly reduced scatter. 
        The dependence on redshift is relatively weak compared to that on S/N.}
    \label{fig:M_mock}
\end{figure*}

To investigate the dependence on data quality, we divide the sample into four 
subsamples based on spectral S/N: S/N $\leq 3$, $3 < \mathrm{S/N} \leq 5$, 
$5 < \mathrm{S/N} \leq 10$, and S/N $> 10$. 
Figure~\ref{fig:M_mock} shows the comparison between the output stellar mass 
($\log M_{\ast, o}$) and the input stellar mass 
($\log M_{\ast, i}$) in six redshift bins spanning $z=0$--0.6. 
It is worth noting that, although the observed wavelength coverage is fixed, 
the corresponding rest-frame wavelength range varies with redshift, 
which may also affect the fitting performance.
Overall, the recovered stellar masses show good agreement with the input values, 
with a clear dependence on S/N. 
For the lowest S/N subsample (S/N $\leq 3$), the stellar masses are systematically 
overestimated, with median offsets of $\sim0.15$--$0.16$ dex and scatter of 
$\sim0.27$--$0.3$ dex across different redshift bins. 
For intermediate S/N ($5 < \mathrm{S/N} \leq 10$), the bias is significantly reduced 
to $\sim0.03$--$0.06$ dex, and the scatter decreases to $\sim0.09$--$0.11$ dex. 
In the highest S/N regime (S/N $> 10$), the bias becomes negligible 
($|\Delta \log M_*| \lesssim 0.015$ dex), and the scatter is further reduced 
to $\sim0.05$ dex.

It is particularly instructive to examine the performance in the extreme low-S/N regime 
(S/N $\leq 3$), where a non-negligible fraction of the DESI BGS spectra reside. 
In this regime, the systematic overestimation of stellar mass becomes significantly 
larger, with median offsets reaching $\sim$0.1--0.2 dex 
and scatter increasing to $\sim$0.2--0.3 dex (see Figure~\ref{fig:M_mock}, top row).
Similarly, the recovered stellar velocity dispersions 
and ages show substantially larger uncertainties. 
While our pipeline yields results for these objects, we caution users that parameters for galaxies with 
S/N $\leq 3$ should be treated with care, as they are dominated by statistical noise 
and may carry significant systematic biases. 
For statistically robust analyses, we recommend applying a threshold of S/N $\gtrsim 5$, 
or properly accounting for the increased uncertainties if including lower-S/N objects.

The systematic overestimation of stellar mass at low S/N can be understood as a 
consequence of degeneracies in spectral fitting, particularly between stellar age, 
dust attenuation, and mass-to-light ratio. 
In low-S/N spectra, the constraints on these parameters become weaker, 
and the fitting tends to favor solutions with slightly higher mass-to-light ratios, 
leading to an overall positive bias in stellar mass. 
This effect is progressively mitigated as the data quality improves.
The dependence on redshift is relatively weak compared to that on S/N, 
although a mild increase in scatter is observed at higher redshift, 
likely reflecting the reduced rest-frame wavelength coverage and lower 
effective S/N. 

Overall, these results demonstrate that our spectral fitting pipeline can 
recover stellar masses with high accuracy for S/N $\gtrsim 5$, 
while providing a quantitative characterization of biases and uncertainties 
in the low-S/N regime typical of DESI data.

\subsection{Recovery of Additional Physical Parameters from Mock Spectra}

In the previous subsection, we quantified the recovery of stellar mass using mock spectra. 
Here, we extend the analysis to additional key physical parameters, including 
stellar velocity dispersion ($\sigma_{\ast}$), dust attenuation ($E(B-V)$), 
and light-weighted stellar age ($t_L$). 
For clarity, instead of presenting full two-dimensional distributions as in 
Figure~\ref{fig:M_mock}, we summarize the results using the median offset 
$\Delta = x_{\mathrm{out}} - x_{\mathrm{in}}$ and the corresponding standard deviation 
in different redshift bins and S/N regimes. The sample is divided into four 
S/N intervals: $\mathrm{S/N} \leq 3$, $3 < \mathrm{S/N} \leq 5$,
$5 < \mathrm{S/N} \leq 10$, and $\mathrm{S/N} > 10$. 
Table~\ref{tab:mock} summarizes the recovery performance for the key parameters 
in a unified format, allowing a direct comparison of their dependence on 
spectral S/N and redshift.
Several important trends can be identified from this table:

(i) Strong dependence on S/N. 
All parameters show a clear improvement in both bias and scatter with increasing S/N. 
For $\mathrm{S/N} > 10$, the median offsets are generally negligible and the scatter is small, 
indicating that the fitting pipeline can reliably recover intrinsic galaxy properties 
under high-quality conditions.

(ii) Systematic biases at low S/N.
At $\mathrm{S/N} \leq 5$, noticeable systematic offsets emerge. 
Stellar mass is mildly overestimated ($\sim 0.1$ dex), while 
light-weighted ages are biased toward younger values. 
These trends reflect well-known degeneracies between age, dust, and metallicity 
in low-S/N spectra.

(iii) Velocity dispersion uncertainties.
The recovery of $\sigma_{\ast}$ exhibits significantly larger scatter compared to other parameters, 
especially at low S/N. This is expected, as velocity dispersion is primarily constrained 
by absorption-line broadening, which becomes increasingly difficult to measure 
when spectral features are noisy or weak.

(iv) Robust recovery of dust attenuation.
The dust attenuation parameter $E(B-V)$ shows minimal systematic bias across all regimes, 
with scatter rapidly decreasing at higher S/N. This indicates that the adopted parametric 
attenuation model remains stable even for relatively noisy spectra.

Overall, these results demonstrate that the reliability of parameter recovery 
is strongly driven by spectral quality. While high-S/N spectra yield accurate and unbiased 
measurements, caution is required when interpreting results from low-S/N data, 
particularly for parameters that depend sensitively on detailed spectral features.

\begin{table*}
\centering
\caption{Recovery statistics from mock spectra for multiple physical parameters. 
Each entry is given as $\Delta / \mathrm{std}$, where $\Delta = x_{\mathrm{out}} - x_{\mathrm{in}}$,
and std is standard deviation. 
Columns correspond to redshift bins ($z=0$--0.6), and rows correspond to different S/N regimes.}
\label{tab:mock}
\begin{tabular}{c|cccccc}
\hline
 & 0$<z\leq$0.1 & 0.1$<z\leq$0.2 & 0.2$<z\leq$0.3 & 0.3$<z\leq$0.4 & 0.4$<z\leq$0.5 & 0.5$<z\leq$0.6 \\
\hline

\multicolumn{7}{c}{\textbf{$\log M_{\ast}$}} \\
\hline
S/N $\leq$3 
& 0.15 / 0.23 & 0.12 / 0.23 & 0.16 / 0.28 & 0.17 / 0.3 & 0.13 / 0.3 & 0.28 / 0.32 \\
3$<\mathrm{S/N}\leq$5 
& 0.12 / 0.16 & 0.09 / 0.14 & 0.11 / 0.16 & 0.11 / 0.17 & 0.1 / 0.17 & 0.12 / 0.18 \\
5$<\mathrm{S/N}\leq$10 
& 0.04 / 0.12 & 0.05 / 0.09 & 0.06 / 0.09 & 0.06 / 0.10 & 0.05 / 0.10 & 0.03 / 0.11 \\
S/N $>$10 
& -0.01 / 0.05 & 0.01 / 0.05 & 0.01 / 0.05 & 0.00 / 0.06 & -0.01 / 0.06 & -0.00 / 0.05 \\

\hline
\multicolumn{7}{c}{\textbf{$\sigma_{\ast}$}} \\
\hline
S/N $\leq$3
& 34.1 / 156.7 & 33.2 / 156.3 & 47.5 / 151.0 & 16.3 / 145.7 & 30.7 / 142.1 & 27.6 / 134.7 \\
3$<\mathrm{S/N}\leq$5
& 14.2 / 41.2 & 12.6 / 54.0 & 14.1 / 52.4 & 10.6 / 51.8 & 11.7 / 56.8 & 8.9 / 57.8 \\
5$<\mathrm{S/N}\leq$10 
& 4.5 / 17.9 & 5.8 / 21.8 & 8.4 / 23.8 & 8.8 / 26.5 & 7.5 / 27.7 & 5.5 / 26.8 \\
S/N $>$10 
& 4.2 / 6.8 & 4.4 / 10.2 & 6.6 / 13.6 & 6.1 / 13.4 & 4.8 / 14.3 & 3.5 / 13.5 \\

\hline
\multicolumn{7}{c}{\textbf{$\log t_L$}} \\
\hline
S/N $\leq$3
& -0.34 / 0.53 & -0.36 / 0.54 & -0.56 / 0.62 & -0.55 / 0.7 & -0.48 / 0.76 & -0.32 / 0.71 \\
3$<\mathrm{S/N}\leq$5
& -0.14 / 0.33 & -0.2 / 0.36 & -0.23 / 0.38 & -0.15 / 0.38 & -0.16 / 0.39 & -0.12 / 0.39 \\
5$<\mathrm{S/N}\leq$10 
& -0.04 / 0.18 & -0.03 / 0.19 & 0.01 / 0.19 & 0.02 / 0.19 & -0.02 / 0.19 & -0.01 / 0.19 \\
S/N $>$10 
& 0.00 / 0.07 & 0.01 / 0.09 & 0.02 / 0.09 & 0.01 / 0.10 & -0.00 / 0.10 & -0.00 / 0.09 \\

\hline
\multicolumn{7}{c}{\textbf{$E(B-V)$}} \\
\hline
S/N $\leq$3
& 0.000 / 0.091 & 0.000 / 0.086 & 0.049 / 0.123 & 0.074 / 0.158 & 0.065 / 0.180 & 0.037 / 0.190 \\
3$<\mathrm{S/N}\leq$5
& 0.000 / 0.044 & 0.004 / 0.044 & 0.017 / 0.053 & 0.015 / 0.059 & 0.024 / 0.069 & 0.026 / 0.083 \\
5$<\mathrm{S/N}\leq$10 
& 0.000 / 0.023 & 0.000 / 0.020 & 0.000 / 0.022 & 0.000 / 0.024 & 0.002 / 0.024 & 0.000 / 0.024 \\
S/N $>$10 
& -0.000 / 0.009 & 0.000 / 0.008 & 0.000 / 0.008 & 0.000 / 0.007 & 0.000 / 0.006 & 0.000 / 0.009 \\

\hline
\end{tabular}
\end{table*}

\subsection{Recovery of Emission-Line Fluxes from Mock Spectra}

\begin{figure}
    \epsscale{1.15}
    \plotone{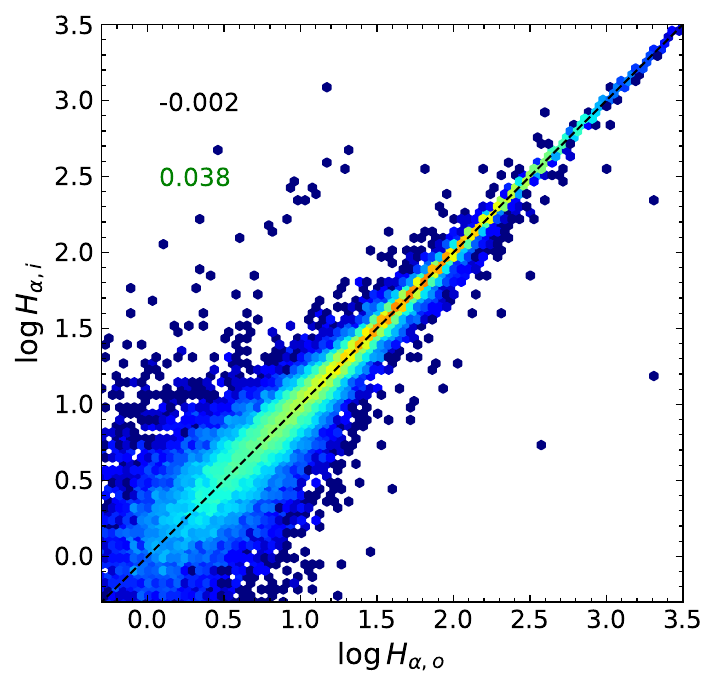}
    \caption{Comparison between input and recovered H$\alpha$ fluxes from mock spectra. 
    The black dashed line indicates the one-to-one relation. 
    The distribution is tightly clustered around this relation, with a median offset 
    of $\Delta \log H\alpha = -0.002$ and a scatter of 0.038 dex.}
    \label{fig:ha_mock}
\end{figure}

In addition to the continuum derived parameters, we also examine the recovery 
of emission-line measurements using the mock spectra. 
Figure~\ref{fig:ha_mock} shows a comparison between the input and recovered 
fluxes of the H$\alpha$ emission line, expressed as 
$\log H_{\alpha, i}$ and $\log H_{\alpha, o}$, respectively.

The recovered H$\alpha$ fluxes show excellent agreement with the input values, 
with nearly all sources lying close to the one-to-one relation. 
The median offset is negligible ($-0.002$ dex), and the scatter is only 0.038 dex, 
demonstrating the high accuracy of our emission-line measurements.
To further validate the reliability of our measurements, 
we also cross-matched our sample with the independent 
DESI DR1 VAC of tellar Mass and Emission Line Catalog
as mentioned in Section~\ref{subsec:aper_corr}.
The comparison yields a median offset of 0.001 dex and 
a scatter of 0.040 dex, which is highly consistent with our mock-based results. 
This excellent agreement confirms that our pipeline is not only 
internally self-consistent but also produces results 
that are robust and consistent with independent measurements.

Here we use H$\alpha$ as a representative example, but similar levels of agreement 
are found for other strong emission lines. Unlike continuum derived parameters, 
the measurement of emission-line fluxes primarily depends on the local signal-to-noise 
ratio of the line itself, rather than the global spectral S/N or the wavelength coverage. 
As a result, we do not subdivide the sample into different S/N or redshift bins 
for this analysis.

Overall, this test confirms that the emission-line fitting procedure is highly robust, 
even under realistic noise conditions, and introduces negligible systematic bias.

\section{Discussion and Summary} \label{sec:diss}

In this work, we have presented a value-added catalog of physical properties 
for 6.7 million galaxies from the DESI DR1 Bright Galaxy Survey, 
derived through full spectral fitting. Given the unprecedented sample size 
and the relatively low S/N of DESI spectra, 
our analysis focuses not only on deriving physical parameters, 
but also on quantifying their reliability using realistic mock spectra.

\subsection{Reliability of spectral fitting at low S/N}

One of the main challenges of DESI spectroscopy is the modest S/N of individual spectra. 
A substantial fraction of galaxies have S/N $\lesssim 5$, 
which significantly limits the constraining power of spectral fitting. 
Our mock tests demonstrate that the accuracy of parameter recovery 
is strongly dependent on S/N. 

For stellar mass, we find a systematic overestimation of $\sim0.1$ dex 
at S/N $\leq 5$, while the bias becomes negligible at S/N $> 10$. 
Similarly, light-weighted stellar ages are biased toward younger values 
in the low-S/N regime, reflecting the well-known degeneracies between 
age, dust attenuation, and metallicity 
\citep[e.g.,][]{Cid-Fernandes_R_2005, Conroy_C_2013}. 
In contrast, emission-line fluxes (e.g., H$\alpha$) are recovered with 
high accuracy and negligible bias, as they depend primarily on the local 
line S/N rather than the global continuum quality.

These results highlight that, although full spectral fitting can be applied 
to large samples of low-S/N spectra, caution is required when interpreting 
continuum-derived parameters. In particular, statistical analyses based on 
low-S/N data should account for systematic biases and increased scatter.

\subsection{Model assumptions and systematic uncertainties}

In this work, we adopt a conventional spectral fitting framework in which 
dust attenuation is described by a parametric attenuation law 
\citep{Calzetti_D_2000}. This choice is motivated by the limited S/N of DESI spectra. 
In previous studies based on high-quality data (e.g., MaNGA), 
more flexible or data-driven approaches have been used to constrain dust attenuation 
independently of stellar population modeling \citep{Li_N_2020, Li_N_2023}. 
However, such methods rely on the ability to separate large-scale continuum 
shapes from small-scale spectral features, which becomes unreliable at low S/N.

It is important to emphasize that our mock spectra are constructed using 
the same modeling assumptions as those adopted in the fitting. 
Therefore, the excellent recovery performance demonstrated in Section~\ref{sec:mock}
primarily reflects the internal consistency of the method, 
rather than capturing all sources of systematic uncertainty. 
In practice, additional biases may arise from model dependencies, 
including the choice of stellar population synthesis models, 
the assumed star formation histories, and the adopted attenuation law. 
Such model-dependent uncertainties have been widely discussed in the literature 
\citep[e.g.,][]{Conroy_C_2013, Siudek_M_2024}, 
and can lead to systematic differences at the level of $\sim0.1$--0.3 dex 
in derived stellar masses.

To quantitatively assess the magnitude of such model-dependent systematics, 
we performed additional tests on a subsample of our mock spectra. 
First, when we fitted the mock spectra (constructed with BC03 SSPs and Calzetti law) 
using the MILES \citep{Vazdekis_A_2010} stellar population library instead of BC03, 
we found a significant systematic bias in the derived stellar masses. 
The bias is about 0.1--0.2 dex even for high-S/N spectra (S/N $>10$) 
and increases to 0.3--0.4 dex in the low-S/N regime (S/N $\leq 3$). 
This demonstrates that the choice of stellar population synthesis models 
is a dominant source of systematic uncertainty, especially for low-quality data. 
In contrast, when we substituted the Calzetti attenuation curve 
with the Milk Way attenuation curve \citep[CCM,][]{Cardelli_JA_1989}, 
the recovered stellar masses remained almost identical to the original 
Calzetti-based fitting results (median offset $\leq 0.01$ dex at S/N $>10$), 
suggesting that the choice between these widely adopted attenuation laws 
is less critical for the BGS dataset. 
Overall, while our internal mock tests demonstrate excellent numerical robustness, 
users of our catalog should be aware that absolute quantities 
such as stellar mass carry an intrinsic systematic uncertainty of 
$\sim$0.1--0.3 dex due to model dependencies.

\begin{figure}
    \epsscale{1.15}
    \plotone{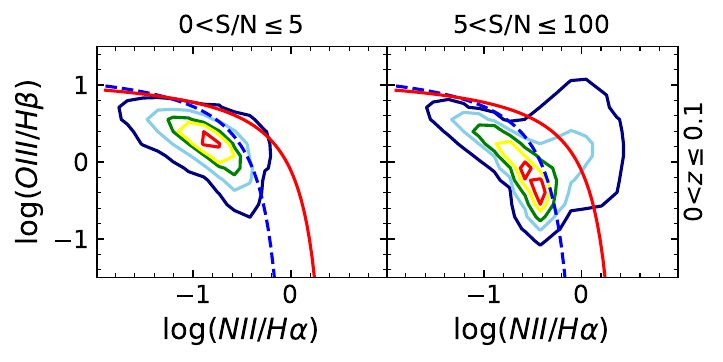}
    \caption{BPT diagnostic diagram for galaxies in the lowest redshift 
             bin (\(z \leq 0.1\)), separated by continuum S/N. 
             The left and right panels show galaxies with \(\mathrm{S/N} \leq 5\) 
             and \(\mathrm{S/N} > 5\), respectively. The meanings of lines and colors
             are similar with those in Figure~\ref{fig:bpt}.}
    \label{fig:bpt_snr}
\end{figure}

\subsection{Implications for statistical studies of galaxy populations}

Despite these limitations, the large statistical power of the DESI BGS sample 
enables robust studies of galaxy populations. 
For quantities such as stellar mass, the combination of moderate individual 
uncertainties and extremely large sample size allows precise measurements 
of population-level trends, such as the stellar mass function and scaling relations. 
However, care must be taken when interpreting subtle features or secondary trends, 
especially in regimes dominated by low-S/N spectra.

The BPT analysis presented in this work illustrates the impact of 
selection effects and observational limitations. 
As demonstrated in Figure~\ref{fig:bpt_snr}, even at a fixed low redshift (\(z \leq 0.1\)), 
galaxies with low and high continuum S/N occupy distinct regions of the BPT diagram. 
Low-S/N galaxies (which are predominantly less massive, as shown in Figure~\ref{fig:hsit_m_t_f}) 
tend to lie on the star-forming sequence, while high-S/N galaxies 
(which are skewed toward higher masses) are shifted toward the composite and AGN regions.
This confirms that the apparent evolution of emission-line ratios with redshift 
is largely driven by the flux-limited nature of the survey 
and the increasing dominance of massive galaxies at higher redshift, 
rather than intrinsic evolution. 
This highlights the importance of forward modeling and careful sample selection 
when using DESI data to study galaxy evolution.

\subsection{Future prospects: physical applications and multi-wavelength extensions}

Beyond methodological improvements, the catalog presented in this work 
opens a wide range of opportunities for studying the co-evolution of stars, gas, and dust in galaxies
which we plan to do.

A key limitation of the current analysis is the use of single-fiber DESI spectra, 
which probe only the central regions of galaxies and are therefore subject to aperture effects. 
Future work will aim to bridge the gap between spatially resolved observations 
and large statistical samples by combining DESI data with integral field spectroscopy surveys 
such as MaNGA. By constructing physically motivated analog samples and modeling aperture biases, 
it will be possible to trace the evolution of dust attenuation properties across cosmic time, 
and to investigate the origin of the two-component attenuation relation between stellar continuum 
and nebular emission \citep{Calzetti_D_2000,Charlot_S_2000,Wild_V_2011,Li_N_2021}.

Another important direction is to extend this catalog through multi-wavelength observations. 
In particular, combining DESI spectroscopy with H\,{\small I} data from FAST surveys \citep{Zhang_CP_2024}
will enable a comprehensive view of the baryon cycle. 
This will allow us to explore scaling relations among gas mass, 
stellar mass, star formation rate, and metallicity, and to quantify the role of gas accretion 
and feedback in shaping galaxy evolution \citep{Lilly_SJ_2013}. 
Such joint datasets will also enable studies of dust growth and destruction processes, 
including the dependence of dust-to-gas ratios on metallicity and star formation activity 
\citep{Remy-Ruyer_A_2014, Galliano_F_2018ARA&A}.
With the statistical power of DESI, this catalog also enables systematic studies of extreme galaxy populations, 
such as extremely metal-poor galaxies and gas-rich quenched systems \citep{Zou_H_2024}. 
These objects provide critical tests of galaxy evolution models in regimes where 
dust shielding, metal cooling, and star formation efficiency may differ significantly 
from those of normal galaxies.

Finally, future work will also explore improved spectral modeling techniques 
for low S/N data, such as stacking the spectra with similar properties. 
In particular, extending non-parametric approaches to dust attenuation 
previously developed for integral field spectroscopy to large spectroscopic samples like DESI 
will help alleviate the degeneracy between stellar age and dust reddening 
\citep{Li_N_2020}, and provide more robust measurements of intrinsic galaxy properties.

Overall, the combination of DESI spectroscopy, multi-wavelength data, 
and advanced modeling techniques will enable a unified framework 
for understanding the interplay between gas, dust, metals, and star formation in galaxies.

\subsection{Summary}

We summarize the main results of this work as follows:

\begin{itemize}

\item We construct a value-added catalog of physical properties 
for 6,662,812 galaxies from the DESI DR1 BGS sample 
using full spectral fitting.

\item The derived stellar masses show good agreement with independent 
photometric estimates, with median offsets $\lesssim 0.1$ dex 
and scatter consistent with previous studies.

\item Mock tests demonstrate that parameter recovery strongly depends on S/N. 
At S/N $\leq 5$, stellar masses are overestimated by $\sim0.1$ dex, 
and stellar ages are biased, while emission-line fluxes remain robust.

\item The spectral fitting pipeline provides reliable measurements 
for S/N $\gtrsim 5$, but caution is required when interpreting 
results in the low-S/N regime.

\item The observed trends in emission-line diagnostics (e.g., BPT diagram) 
are primarily driven by selection effects and observational biases, 
rather than intrinsic galaxy evolution.

\item This catalog provides a valuable resource for statistical studies 
of galaxy populations in the low-redshift universe, 
and serves as a foundation for future analyses based on stacked spectra 
and multi-wavelength data.

\end{itemize}

Overall, this work demonstrates that, despite the challenges posed by low-S/N data, 
full spectral fitting of DESI spectra can yield robust constraints on galaxy 
physical properties when combined with careful validation and statistical analysis. 
The value-added catalog presented in this work is publicly available at 
Science Data Bank with the DOI: \url{http://www.doi.org/10.57760/sciencedb.35360}.

\begin{acknowledgments}
The authors acknowledge the supports from National Key R\&D Program of China 
(grant Nos. 2025YFE0202300, and 2022YFA1602902), 
the National Natural Science Foundation of China 
(NSFC; grant Nos. 12120101003, 12373010, 12233008, and 12503019) 
and China Manned Space Project (No. CMS-CSST-2025-A06). 
The authors also acknowledge the National Key R\&D Program of China 
(grant Nos. 2023YFA1607804, 2023YFA1608100, 2023YFA1607800), 
the Strategic Priority Research Program of the Chinese Academy of Sciences 
with Grant Nos. XDB0550100 and XDB0550000 and the Programs of National Astronomical 
Observatories Chinese Academy of Sciences with Grant Nos. E5ZQ7801, E5ZB7801, and E4TG2001.
The authors thank Cheng Li for his guidance and suggestions 
during the early stages of this work.
\end{acknowledgments}


\software{Astropy \citep{2013A&A...558A..33A,2018AJ....156..123A,2022ApJ...935..167A},
          CIGALE \citep{Boquien_M_2019}, FastSpecFit \citep{Moustakas_J_2023},
          Redrock \citep{Brodzeller_A_2023}
          }

\appendix
\section{Catalog Description} \label{app:data_model}
In this work, we make use of two main data products: 
(i) the official DESI DR1 spectroscopic catalog, and 
(ii) the value-added catalog of physical properties derived in this work.
The catalogs also can be downloaded from Science Data Bank \footnote{\url{https://www.scidb.cn/}},
with the link \footnote{\url{http://www.doi.org/10.57760/sciencedb.35360}}.

\subsection{DESI DR1 Spectroscopic Catalog}

The primary data source is the DESI DR1 redshift catalog 
\texttt{zall-pix-iron-bgs.fits.gz}, which is a subset of the full DESI catalog 
\texttt{zall-pix-iron.fits} restricted to BGS targets. 
All columns are identical to the official DESI release.
The detailed data model of this catalog is publicly available online%
\footnote{\url{https://desidatamodel.readthedocs.io/en/latest/DESI_SPECTRO_REDUX/SPECPROD/zcatalog/v1/zall-pix-SPECPROD.html}}.
Here we list the main parameters in Table~\ref{tab:zcat}
which are used in this work.

\begin{table*}
\centering
\caption{Selected columns from the DESI DR1 BGS catalog.}
\label{tab:zcat}
\begin{tabular}{lll|lll}
\hline
Column & Unit & Description & Column & Unit & Description \\
\hline
TARGETID & -- & Unique DESI target ID 
& Z & -- & Redshift from Redrock \\

SURVEY & -- & Survey name 
& ZERR & -- & Redshift uncertainty \\

PROGRAM & -- & Observing program 
& ZWARN & -- & Redshift warning flag \\

HEALPIX & -- & HEALPix index 
& EBV & mag & Galactic extinction \\

FLUX\_G & nanomaggy & $g$-band flux 
& FLUX\_R & nanomaggy & $r$-band flux \\

DESI\_TARGET & -- & Target bitmask 
& BGS\_TARGET & -- & BGS selection flag \\

MWS\_TARGET & -- & MWS selection flag 
& SCND\_TARGET & -- & Secondary targets \\
\hline
\end{tabular}
\end{table*}

\subsection{Value-added Catalog from Spectral Fitting}

The value-added catalog generated in this work is stored in the file 
\texttt{BGS\_catalog.fits.gz}. 
It contains the spectral fitting results for 7,033,357 spectra, organized into three extensions:

\begin{itemize}
    \item \texttt{FIT\_RES}: stellar continuum fitting results and derived physical parameters
    \item \texttt{FIT\_EMLINES}: emission-line measurements
    \item \texttt{ZCAT}: basic identifiers linking to the DESI catalog
\end{itemize}

\subsubsection{FIT\_RES: Stellar population and continuum fitting}
This extension contains the columns describing the spectral fitting results. 
Table~\ref{tab:fitres} summarizes the main parameters.

\begin{table*}
\centering
\caption{Columns in the \texttt{FIT\_RES} extension.}
\label{tab:fitres}
\begin{tabular}{lll|lll}
\hline
Column & Unit & Description & Column & Unit & Description \\
\hline
snr & -- & Signal-to-noise ratio 
& nor & -- & Normalization factor \\

ve & km\,s$^{-1}$ & Stellar velocity 
& vd & km\,s$^{-1}$ & Velocity dispersion \\

weights & -- & SSP weights (150 templates) 
& ebv & mag & Dust attenuation $E(B-V)$ \\

dof & -- & Degrees of freedom 
& chi2 & -- & $\chi^2$ of fit \\

abmag\_g & mag & Synthetic $g$ magnitude 
& abmag\_r & mag & Synthetic $r$ magnitude \\

M & $M_\odot$ & Stellar mass 
& age\_L & log(yr) & Light-weighted age \\

z\_L & -- & Light-weighted metallicity 
& age\_M & log(yr) & Mass-weighted age \\

z\_M & -- & Mass-weighted metallicity 
& age\_L2 & log(yr) & Alternative light-weighted age \\

z\_L2 & -- & Alternative metallicity 
& age\_M2 & log(yr) & Alternative mass-weighted age \\

z\_M2 & -- & Alternative metallicity 
& EW\_hd\_a & \AA & H$\delta_A$ index \\

D4000 & -- & 4000\AA\ break 
& D4000\_n & -- & Narrow D4000 \\

Mgb & \AA & Magnesium index 
& Fe5270 & \AA & Iron index \\

Fe5335 & \AA & Iron index 
& -- & -- & -- \\
\hline
\end{tabular}
\end{table*}

\subsubsection{FIT\_EMLINES: Emission Line Measurements}
This extension contains measurements for 18 emission lines, 
including [O\,II], H$\beta$, [O\,III], H$\alpha$, [N\,II], and [S\,II],
which can be seen in Table~\ref{tab:emlines}.

Each emission line is characterized by eight quantities: 
the first four correspond to the line flux, equivalent width, center wavelength, 
and Gaussian-fitted sigma, while the latter four are the respective errors.

\begin{table*}
\centering
\caption{Emission lines included in the catalog.}
\label{tab:emlines}
\begin{tabular}{lll|lll}
\hline
Line & Number & Description & Line & Number & Description \\
\hline
OII3726 & 0 & [OII] doublet 
& OII3729 & 1 & [OII] doublet \\

NeIII3869 & 2 & Neon line 
& H$\delta$ & 3 & Balmer line \\

H$\gamma$ & 4 & Balmer line 
& OIII4363 & 5 & Auroral line \\

H$\beta$ & 6 & Balmer line 
& OIII4959 & 7 & Oxygen line \\

OIII5007 & 8 & Oxygen line 
& HeI5876 & 9 & Helium line \\

NaD5891 & 10 & Sodium absorption 
& OI6300 & 11 & Neutral oxygen \\

NII6548 & 12 & Nitrogen line 
& H$\alpha$ & 13 & Balmer line \\

NII6583 & 14 & Nitrogen line 
& SII6717 & 15 & Sulfur line \\

SII6731 & 16 & Sulfur line 
& ArIII7135 & 17 & Argon line \\
\hline
\end{tabular}
\end{table*}

\subsubsection{ZCAT: Cross-match Information}
The \texttt{ZCAT} extension provides identifiers that link each entry in the 
value-added catalog to the original DESI catalog
which can be seen in Table~\ref{tab:zcat_bgs}.
\begin{table*}
\centering
\caption{Columns in the \texttt{ZCAT} extension.}
\label{tab:zcat_bgs}
\begin{tabular}{ll}
\hline
Column &  Description \\
\hline
TARGETID &  Unique DESI target ID \\
SURVEY &  Survey name \\
PROGRAM &  Observing program \\
HEALPIX & HEALPix index \\
\hline
\end{tabular}
\end{table*}

Since the BGS sample is selected directly from the DESI DR1 catalog, 
the ordering of sources in this extension is consistent with the input catalog, 
allowing direct cross-matching without ambiguity.

\bibliography{ref}{}
\bibliographystyle{aasjournalv7}

\end{CJK*}
\end{document}